\documentclass[aps,pra,twocolumn,a4paper,superscriptaddress]{revtex4} 
\usepackage{amsmath}
\usepackage{amsfonts}
\usepackage{amssymb}
\usepackage{epsfig}
\usepackage{graphicx}
\usepackage{natbib}
\usepackage{doi}
\usepackage{hyperref}

\usepackage{xcolor} 

\usepackage{xfrac}

\begin{document}

\title{Theoretical study of stimulated and spontaneous Hawking effects \\ 
from an acoustic black hole in a hydrodynamically flowing fluid of light
}
\author{Pjotrs Gri\v{s}ins}
\email{pjotrs.grisins@unige.ch}
\affiliation{DQMP, University of Geneva, 24 Quai Ernest Ansermet, 1211 Geneva, Switzerland}
\affiliation{INO-CNR BEC Center and Universit\`a di Trento, Via Sommarive 14, 38123 Povo, Italy}

\author{Hai Son Nguyen}
\affiliation{Institut des Nanotechnologies de Lyon, 36 Avenue Guy de Collongue, 69130 \'Ecully, France}

\author{Jacqueline Bloch}
\affiliation{Centre de Nanosciences et de Nanotechnologies, CNRS, Univ. Paris-Sud, Universit\'e Paris-Saclay, C2N -- Marcoussis, 91460 Marcoussis, France}


\affiliation{Physics Department, \'Ecole Polytechnique, F-91128 Palaiseau Cedex, France}

\author{Alberto Amo}
\affiliation{Centre de Nanosciences et de Nanotechnologies, CNRS, Univ. Paris-Sud, Universit\'e Paris-Saclay, C2N -- Marcoussis, 91460 Marcoussis, France}


\author{Iacopo Carusotto}
\affiliation{INO-CNR BEC Center and Universit\`a di Trento, Via Sommarive 14, 38123 Povo, Italy}

    \begin{abstract} \noindent
We propose an experiment to detect and characterize the analog Hawking radiation in an analog model of gravity consisting of a flowing exciton-polariton condensate. Under a suitably designed coherent pump configuration, the condensate features an acoustic event horizon for sound waves that at the semiclassical level is equivalent to an astrophysical black hole horizon. We show that a continuous-wave pump-and-probe spectroscopy experiment allows to measure the analog Hawking temperature from the dependence of the stimulated Hawking effect on the pump-probe detuning. We anticipate the appearance of an emergent resonant cavity for sound waves between the pump beam and the horizon, which results in marked oscillations on top of an overall exponential frequency dependence. We finally analyze the spatial correlation function of density fluctuations and identify the hallmark features of the correlated pairs of Bogoliubov excitations created by the spontaneous Hawking process, as well as novel signatures characterizing the emergent cavity.   
		\end{abstract}

    \maketitle

\section{Introduction}\label{s:intro}

Since its first prediction in 1975 \cite{Hawking1975}, Hawking radiation from astrophysical black holes remains one of most intriguing phenomena in modern physics, as it questions fundamental aspects of our understanding of the physical world down to the unitarity of quantum mechanics \cite{Hawking2005,Almheiri2013}. Unfortunately, a direct experimental detection of Hawking radiation is made difficult by its extremely low temperature, which is expected to be on the few tens of nK range for solar mass black holes.

As it was first proposed by Unruh \cite{Unruh1981}, it is nowadays possible to study the Hawking effect in tabletop experiments based on the so-called analog model idea, where the propagation of low-energy excitations on top of an inhomogeneous and moving medium can be reformulated in terms of a quantum field propagating on a curved {space-time~\cite{Birrell1982}. As a direct consequence, one can anticipate that the same Hawking mechanism that is expected to be active in gravitational black holes will be responsible for a thermal emission of excitations from analog black hole horizons~\cite{Barcelo2011}.
In condensed-matter terms,} this effect can be interpreted as the inhomogeneous flow converting the zero-point vacuum fluctuations into an observable two-mode squeezed vacuum state with a thermal spectrum. In addition to their intrinsic interest, analog models may shed light on the long-standing questions in cosmology, such as the trans-planckian problem \cite{Brout2007,Barcelo2011}, the effects of possible breakdown of the Lorentz-invariance at high energies \cite{Unruh1995,Finazzi2012}, and {the back-action of the produced particles on the background metric~\cite{BalbinotBR1,BalbinotBR2}.}


The original proposal \cite{Unruh1981} suggested to use water as the background medium, and sound waves as the low-energy fluctuations. While intriguing experiments using surface waves in water were reported~\cite{Weinfurtner2011} and debated in the literature~\cite{Michel2014,Euve2015}, the estimated temperature of the spontaneous Hawking radiation for any realistic flow remains however orders of magnitude too low to make any quantum effects observable before actual freezing. Other systems hold a stronger promise as analog models to study quantum features, e.g. dilute superfluids such as ultracold Bose-Einstein condensed (BEC) atomic gases, and nonlinear optical systems.

Ultracold atomic gases were first proposed as analog models in~\cite{Garay2000} and theoretically explored in great detail in the last decade~\cite{Balbinot2008a,Carusotto2008b,Macher2009,Recati2009,Zapata2011,DeNova2012,Finazzi2013}. From the experimental point of view, the creation of an acoustic black hole horizon was first reported in 2010: the supersonic flow was generated by imposing an engineered moving step-like potential to a harmonically trapped condensate of $^{87}$Rb atoms~\cite{Lahav2010}. A few years later, the same group reported the observation of self-amplifying Hawking emission via the analog black hole laser instability~\cite{Steinhauer2014}.

In non-linear optical systems the first steps in the direction of studying analog black hole were reported in \cite{Philbin2008a}. A solitonic pulse of light propagating in a Kerr nonlinear medium creates regions of different index of refraction and hence different speed of light. For suitably chosen pulse parameters, this solitonic boundary may create a moving `event horizon' that can be crossed by light in one direction only. The intriguing experimental observations reported in \cite{Belgiorno2010} were followed by a vivid discussion on the interpretation of the observed phenomena \cite{Schutzhold2011,Belgiorno2011,Liberati2012,Finazzi2014}.


%
 
In the past few years novel non-linear optical configurations started attracting a great interest from the community, the so-called quantum fluids of light \cite{Carusotto2013}{. After a series of pioneering experiments demonstrating superfluidity and superfluid hydrodynamic behaviours~\cite{Amo2009,Amo2011,Nardin2011}, it was soon proposed that these novel quantum many-body systems are also able to host analog black hole horizons \cite{Marino2008,Marino2009a,Fouxon2009,Solnyshkov2011,Gerace2012}.}
Recently the formation of an analog black hole in such systems was first reported using a semiconductor microcavity device in the strong light-matter coupling regime~\cite{Nguyen2015}.
In that work, the microcavity was laterally patterned to form a 1D channel for polaritons containing a localised potential defect (see Fig.~\ref{f:setup}). A suitable coherent pump was used to inject a fluid of interacting exciton-polaritons into the channel: {for sufficiently strong pump intensities, a black hole horizon for density waves on top of the fluid spontaneously appeared at the position of the defect.}

After having created an horizon and fully characterized its properties, the next big challenge will be to detect the analog Hawking radiation resulting from quantum fluctuations. To this purpose, Ref.~\cite{Gerace2012} proposed to measure the intensity correlations on both sides of the horizon to detect the spontaneous Hawking emission of phonons. However, the weakness of the expected signal makes the experiment extremely challenging as the thermal Hawking phonons may be overshadowed by noise of different physical origins, e.g. light scattering on device imperfections, incoherent luminescence from the semiconducting material, a non-zero temperature of the fluid of light due to interaction with phonons, shot-noise in the optical measurement. In addition to this, the intrinsically driven-dissipative nature of the photon/polariton fluid is responsible for an additional source of quasi-thermal noise in the non-equilibrium stationary state, with an effective temperature typically of the order of the interaction energy of the fluid \cite{Busch2014,Grisins2014b}. Also from the theoretical point of view, the intrinsically driven-dissipative nature of the polariton fluid forbids a direct use of the nowadays well-known theory {of analog Hawking effect in} atomic condensates and raises questions on very fundamental aspects of analog Hawking radiation, including its thermal character and its quantum correlation properties.

%
%
In the present work we propose and characterize different experimental strategies to enhance and measure the analog Hawking radiation from a black hole horizon in a one-dimensional fluid of light. In particular, we analyze the thermal character of the expected Hawking emission in an experimentally realistic situation. While our discussion is focussed on the configuration used in the recent experiment using exciton-polaritons in a semiconductor microcavity~\cite{Nguyen2015}, most of our conclusions directly transfer to other material platforms, for instance generic fluids of cavity photons interacting via a $\chi^{(3)}$ optical nonlinearity~\cite{Carusotto2013}.


The paper is organized as follows. In Section~\ref{s:analog} we review the experimental setup of \cite{Nguyen2015}, which serves as a foundation for the experiment we are proposing in this work. Section~\ref{s:stim} is dedicated to the stimulated Hawking effect, where we show that by sending a weak probe pulse to the horizon and analyzing the reflected signal it is possible to measure the Hawking temperature. This idea was first discussed in the context of atomic condensates~\cite{Macher2009,Recati2009} and surface waves on water~\cite{Rousseaux,Weinfurtner2011}{, and then transferred to fluids of light in~\cite{Gerace2012}. While the optical set-up proposed in that work requires a very sophisticated time-resolved apparatus for the generation and detection of short probe wavepackets, here we focus on a continuous-wave scheme that} dramatically reduces the experimental difficulty of extracting quantitative information on the {\em stimulated Hawking processes} during scattering on the horizon. While a clear signature of the thermal Hawking spectrum with a temperature in the Kelvin range is found in the envelope of the spectrum, the emergent Fabry-Perot cavity for sound waves that naturally forms between the pump and the horizon introduces marked oscillations on the Hawking spectrum.

In Section~\ref{s:spon} we analyze the {\em spontaneous Hawking effect} and we provide {a thorough} discussion of the different features that {appear} in the spatial correlation function of the intensity noise of the microcavity emission: {clear signatures of the spontaneous Hawking effect are found in the usual off-diagonal correlations on either side of the horizon and new correlation features are pointed out in the inner super-sonic region and physically explained in terms of the frequency-modulation of the Hawking emission due to the emergent resonant cavity.} Section~\ref{s:concl} finally summarizes our findings.

\section{The physical system}
\label{s:analog}

\begin{figure}
\includegraphics[width=0.85\columnwidth,clip]{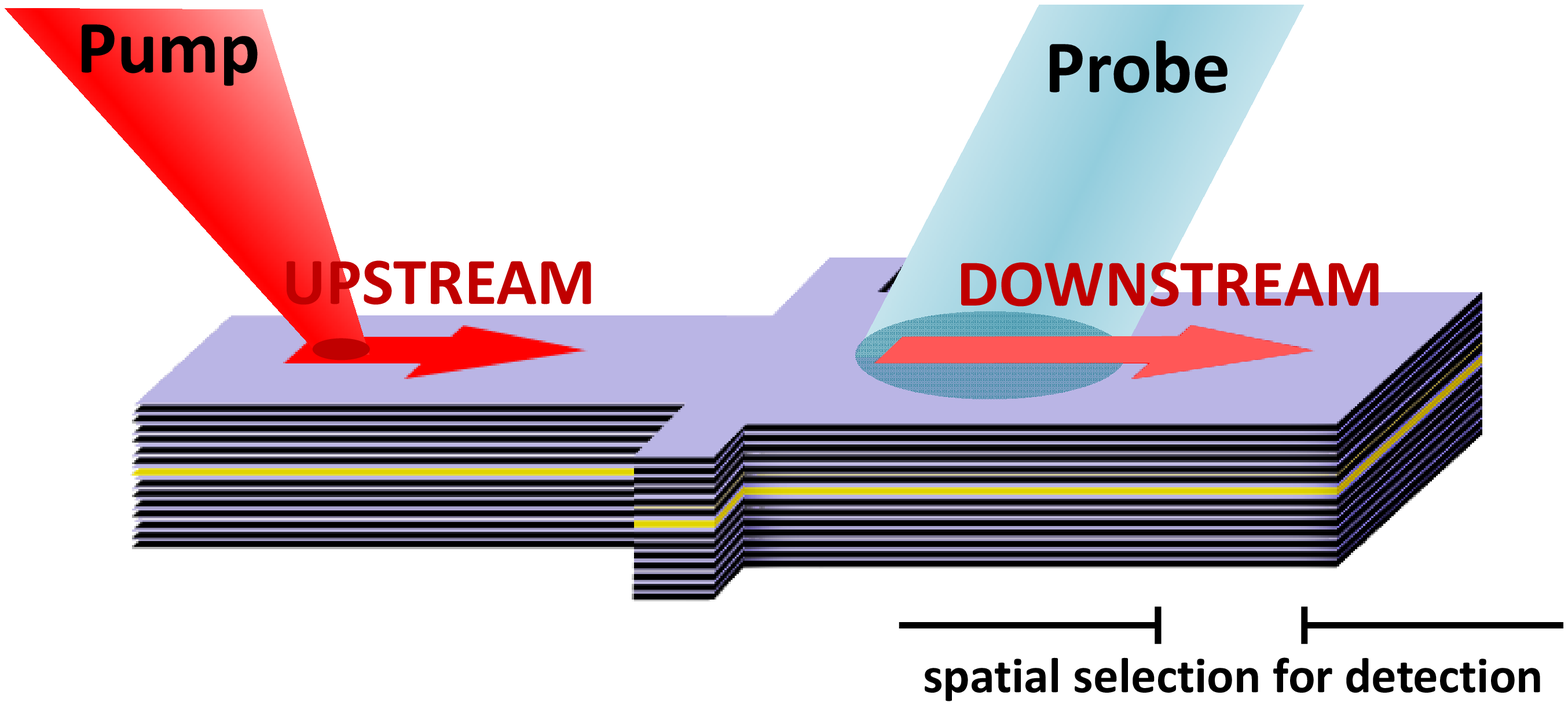}\\
\vspace*{-1.5cm}
\includegraphics[width=0.85\columnwidth,clip]{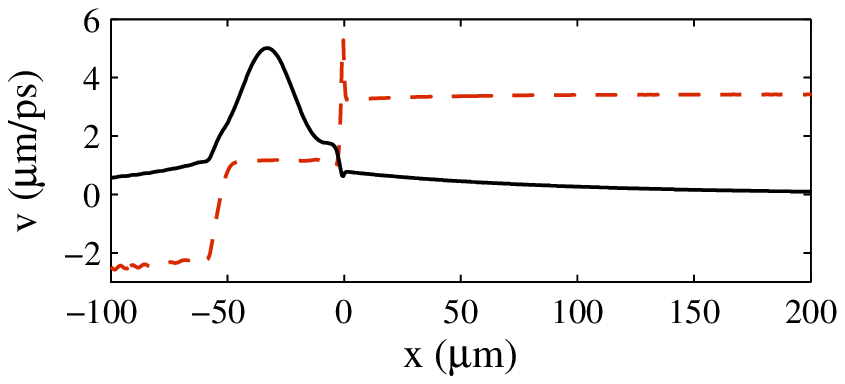}
\caption{
{\bf Upper panel:} Setup of the proposed experimental device to observe the stimulated Hawking radiation. It consists of a laterally patterned semiconductor microcavity device in the form of a {long one-dimensional photonic wire oriented} along the $x$ direction, and with a {micron-sized} $x$-dependent lateral profile designed in a way to facilitate observation of analog Hawking effects.
The flowing condensate is created by a focussed coherent pump laser beam incident at a finite angle on the device. The attractive defect potential created by the wider region around $x=0$ creates an analog black hole horizon in its vicinity, which separates an upstream region of subsonic flow from a downstream region of supersonic flow. A weak probe laser is shone in the downstream region so to generate the Bogoliubov waves which then stimulate the Hawking radiation. The cavity emission is collected in the far downstream region by applying a spatial mask to filter out the strong pump and probe signal. 
{\bf Lower panel:} spatial profile of the polariton flow velocity and of the local speed of sound along the wire. The acoustic event horizon {is located where} the flow velocity starts exceeding the speed of sound. The high-density region around $x=-30\ \mu$m corresponds to the spatial position of the pump beam. 
}
\label{f:setup}
\end{figure}


The proposed experimental setup is sketched in Fig.~\ref{f:setup}. The essential parts of the setup are the same as in the experiment \cite{Nguyen2015} where a black hole configuration in a flowing polariton fluid was first observed. 

The physical device consists of a GaAs microcavity sandwiched between a pair of planar Bragg reflectors with alternating $\lambda/4$ layers of GaAs/AlGaAs. An InGaAs quantum well is inserted in the microcavity layer, whose thickness is chosen so that the cavity mode is resonant with the quantum well excitonic transition. In this way, the resulting excitation modes have a mixed light-matter character and go under the name of {\em exciton-polaritons}. Thanks to their excitonic component, polaritons show strong binary interactions that lead to collective behaviours in the polariton fluid. The photonic component of the polariton allows for an easy generation and manipulation of the polariton fluid using incident light beams, as well as for a straightforward diagnostic of the fluid dynamics from the emitted light. The interested reader can find a general introduction to the general physics of quantum fluids of polaritons in microcavity devices in the recent review~\cite{Carusotto2013}.

In order to keep the fluid dynamics the simplest, the cavity is etched to form a {$500\,\mu$m} long polaritonic wire with a $3\,\mu$m transverse width where the polariton dynamics is effectively one-dimensional. As it is sketched in the upper panel of Fig.~\ref{f:setup}, polaritons are injected into the cavity by spatially focussing a continuous wave coherent {\em pump} laser with a finite incidence angle with respect to the normal. This generates a stationary one-dimensional flow of polaritons that propagate along the wire and eventually hit an engineered defect. The defect consists in a localized {$1\,\mu$m long} broadening of the wire {to a width of $5.6\,\mu$m} that can be modeled as a localized attractive potential well for polaritons. 

While at low injected powers polaritons partially scatter from the defect, at sufficiently strong pump powers superfluid effects set in and a black hole acoustic event horizon is spontaneously formed in the flow profile in the vicinity of the defect. The horizon separates an upstream region of sub-sonic flow from a downstream region of super-sonic flow. In the gravitational analogy, these regions are analogous to the exterior and the interior of the black hole, respectively. 
A sketch of the spatial profile of the flow and sound speeds is displayed in the {lower} panel of Fig.\ref{f:setup}.

As the only addition with respect to the scheme used in Ref.~\cite{Nguyen2015}, the attractive defect is here supplemented by a homogeneous attractive potential in the downstream region: in this way, the flow velocity in the downstream region is increased, which leads to a reinforced surface gravity at the horizon and therefore to a higher Hawking temperature. Furthermore, as we will see below, the higher flow speed facilitates separation of the different momentum components in the pump-and-probe experiment and therefore the identification of the Hawking signal. As sketched in the upper panel of Fig.~\ref{f:setup}, in future experiments it will be straightforward to include this attractive potential {by means of a slight increase of the transverse size of the wire in the downstream region on the order of $1\,\mu$m.}

One of the main advantages of microcavity polaritons with respect to other systems is the easy access to observables such as the polariton density and momentum. Diagnostics of the polariton fluids is made via the emitted light which inherits the statistical properties of the in-cavity field~\cite{Carusotto2013}. For instance, the real space intensity profile of the microcavity emission reproduces the spatial density distribution of the polaritons and the correlation function of the intensity noise provides direct information on the corresponding density fluctuations, which are expected to contain information on the Hawking radiation~\cite{Gerace2012}. Furthermore, the momentum distribution of the fluid can be extracted from the angular distribution of the emission. In particular, a spatially-resolved momentum distribution can be obtained as in the experiment~\cite{Nguyen2015} by first spatially filtering the emission in the real space and {then observing the emission in the far field.}.

A second advantage of polariton fluids is the possibility to directly probe and manipulate the state of the fluid using additional lasers. For instance, the spectrum of excitations {on top of the flowing condensate} can be probed using a weak continuous-wave laser beam. The polaritons injected by this {\em probe} beam are strongly dressed {by the nonlinear interaction with the condensate} and propagate as collective Bogoliubov excitations in the fluid. The wavevector and frequency of these Bogoliubov excitations can be tuned by varying the frequency and the incidence angle of the probe beams. 

{In contrast to gravitational systems, in our analog model choosing a large probe incidence angle with an opposite sign with respect to the pump beam allows for the generation of large wavevector single-particle-like Bogoliubov excitations that propagate in the upstream direction against the horizon. A configuration of this kind appears to be most favourable for the study of stimulated Hawking processes that is reported in the next} Sec.~\ref{s:stim}.

\section{Stimulated Hawking radiation}\label{s:stim}

\subsection{Theoretical model and general considerations}

We work with the exciton-polaritons of the lower polariton branch close to the bottom of the dispersion relation, which at the mean-field level are described by the driven-dissipative Gross-Pitaevskii equation with a standard parabolic single-particle dispersion \cite{Carusotto2013}
\begin{multline}
\label{eq:mf_gpe}
i\frac{d}{dt}  \phi(x,t) = \\ =\left[ -\frac{\hbar}{2m} \,\frac{d^2}{dx^2}+ V(x) + g |\phi (x,t) |^2 - i \frac{\gamma }{2} \right] \phi (x,t) +
	\\ 
	 +F_{p}(x,t)+F_{s}(x,t),
\end{multline}
where the effective polariton mass along the wire is taken to be {$m/\hbar^2= 0.26\,\textrm{meV}^{-1}\,\mu\textrm{m}^{-2}$}, the one-dimensional polariton-polariton interaction constant $g$ is assumed to be constant along the wire with a value such that $\hbar g = 0.005\,\textrm{meV}\,\mu$m~\footnote{We have verified that the physics remains qualitatively unchanged if a spatially varying interaction strength $g$ following the wire profile is used in the simulations.}
, 
and the polariton decay rate  $\gamma$ is assumed for simplicity to be position- and momentum-independent with a value such that $\hbar \gamma= 0.047\,$meV. $V(x)$ is the effective potential experienced by polaritons as a result of the transverse confinement and is roughly related to the transverse width $w(x)$ by
\begin{equation}
 V(x)\approx \frac{\hbar^2 \pi^2}{2m\,w(x)^2}.
\end{equation}
In our calculations, we assumed the potential to have the model form
\begin{equation}
 V(x)=V_{\rm attr}\,\Theta(x)+V_{\rm def}\,e^{-x^2/w_{\rm def}^2}
\end{equation}
with a defect potential strength of $V_{\rm def}=-0.85\,$meV localized in a region of size $w_{\rm def}=0.75\,\mu$m and an attractive potential of strength $V_{\rm attr}=-0.4\,$meV in the downstream region .

Within the parabolic band approximation, the fluid velocity $v(x,t)$ and the speed of sound $c(x,t)$ are defined in terms of the field $\phi(x,t)$ as 
\begin{eqnarray} 
v(x,t)=\mathrm{Im}(\phi^* \partial_x \phi)/m \\ 
c(x,t)=\sqrt{g |\phi(x,t)|^2/m}.
\end{eqnarray}

The profile of the pump beam is taken to have a carrier frequency $\omega_{p}$, an {in-plane} carrier wavevector $k_{p}$, and a slow envelope along the wire of Gaussian form of width $\sigma_{p}$,
\begin{equation}
F_{p}(x,t)=
F^0_{p}\,e^{-\log(2)\,(x-x_{p})^2/(\sigma_{p}/2)^2}\,e^{i(k_{p} x - \omega_{p} t)}.
\end{equation}
In our calculations, the pump {profile is inspired to} the experimental parameters of~\cite{Nguyen2015}, with a width $\sigma_p = 17\,\mu$m, and centered at {$x_p=-33\,\mu$m} upstream from the defect. Polaritons then propagate from the pump spot towards the horizon. Due to their finite lifetime, in the steady state the polariton density shows a spatially decaying profile, resulting in a space dependent velocity of sound.

\begin{figure*}
\includegraphics[width=0.95\textwidth]{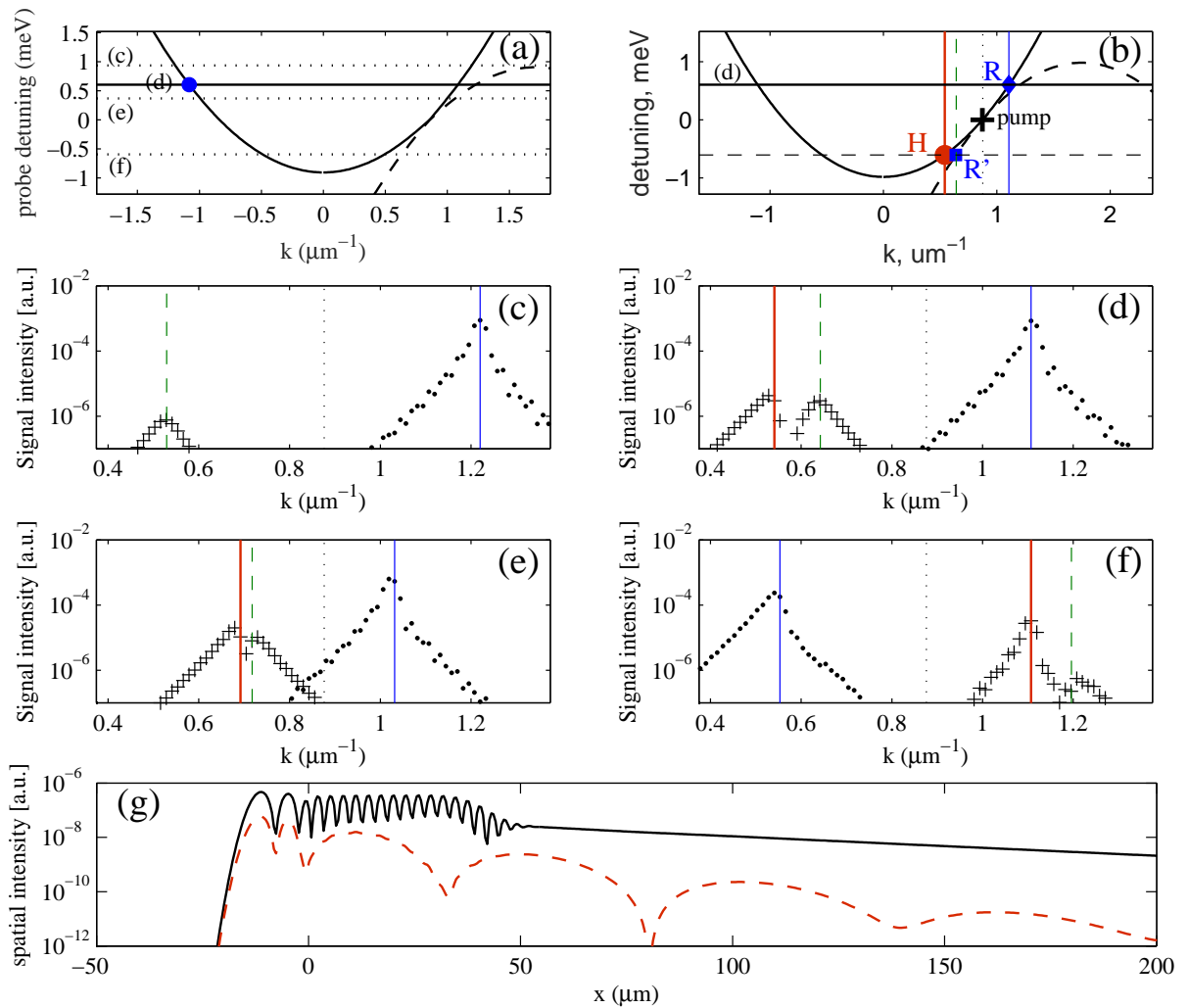}
\caption{
\textbf{Top-left (a) panel:} Dispersion relation of the Bogoliubov quasiparticles at the probe beam position ($x \approx 40\ \mu$m). The vertical scale is given by the probe detuning $\hbar \Delta = \hbar \omega_{s} - \hbar \omega_{p}$: a positive detuning means that the probe is blue detuned relative to the pump. \textbf{Top-right (b) panel:} The same Bogoliubov dispersion relation in the signal collection region ($x \approx 160\ \mu$m) {where the condensate density is even lower}.
\textbf{Central (c-f) panels:} Examples of angular distribution of the collected emission for the different values of the probe detuning $\hbar \Delta \approx 0.94, 0.60, 0.37, -0.60$ meV indicated by the horizontal lines in {panel (a)}. The dots refer to the emission at the probe frequency $\omega_s$, while the crosses refer to the emission at the four-wave mixed frequency $\omega_4=2\omega_{p} - \omega_{s}$. The vertical lines in (c-f) indicate the prediction of the Bogoliubov theory for the wavevectors of outgoing modes. For the $\hbar \Delta \approx 0.60$ meV case of panel (d), these modes are also indicated by the vertical lines in panel (b): the label H and the thick red solid line indicate the Hawking signal, the label $R$ and the blue thin solid line indicate the reflected component, and the label $R'$ and the dashed line indicate the four-wave-mixed partner of the reflected component. Note that for $\hbar\Delta = 0.94$ meV there is no visible Hawking signal. The blue dot in panel (a) indicates the ingoing Bogoliubov mode that is resonantly injected by the probe beam. 
\textbf{Bottom (g) panel:} spatial profile of the emission filtered at the probe frequency $\omega_{s}$ (solid black line) and at the four-wave mixed frequency $\omega_4$ (solid red line). The probe detuning is $\hbar \Delta \approx 0.60$ meV and corresponds to panel (d). The spatial beats in the $\omega_{4}$ signal {in the inner $x>0$ region} correspond to the interference between the two peaks of comparable magnitude {that are indicated as the $H$ and the $R'$ peaks on} the angular distribution of the $\omega_4$ emission shown in panel (d). The {faster} spatial beats in the $\omega_{s}$ signal {in the $0\,\mu\textrm{m}<x<x_s$ region between the probe and the horizon are due to interference of the incident and the reflected $R$ waves. Finally, the beats that are visible at both $\omega_{s,4}$ frequencies in the $-10\,\mu\textrm{m}<x<0\,\mu\textrm{m}$ region right in front of the horizon are a signature of the emergent cavity between the pumped region and the horizon}.  
}\label{f:peaks}
\end{figure*}

An analogous Gaussian profile centered at $x_s=40\,\mu$m downstream of the defect and of approximate {full} width $\sigma_s=7\,\mu$m is taken for the probe beam of frequency $\omega_s$, slightly detuned by $\Delta=\omega_s-\omega_p$ from the pump. Its {in-plane wavevector} $k_s$ is tuned on resonance with the Bogoliubov spectrum of excitations of the fluid at the position of the probe, as indicated by the points in Fig.\ref{f:peaks}(a). The probe amplitude $F^0_s$ is chosen small enough not to exceed the regime of validity of the linearized Bogoliubov theory. The Bogoliubov excitations created by the probe on top of the fluid will then propagate away from the probe spot at a speed determined by the group velocity of the Bogoliubov dispersion, for the chosen parameters in the upstream direction towards the horizon. 

It is worth noting that for all choices of parameters considered in this work, all other Bogoliubov branches are suppressed as their wavevector and/or their frequency are far from resonance for both direct and four-wave mixing~\cite{Wouters:PRB2009b} processes. In the diagram of Fig.\ref{f:peaks}(a), the pole of a four-wave mixing resonance occurs when the probe frequency and wavevector lie in the vicinity of the dashed line.

As it is sketched in the upper panel of Fig.\ref{f:setup}, the Bogoliubov waves that emerge from scattering of the probe off the horizon are detected from the emission collected in a finite spatial window in the downstream region~\cite{Nguyen2015}. To optimize the analysis, the collection region is located further downstream than the probe so to avoid contribution of the probe waves incident on the horizon. Additionally, the collection region is chosen to have a wide extension so to guarantee a good momentum selectivity.
Specifically, the filter function used in the calculations has the form
\begin{multline}
F(x)=\frac{1}{4} \left[1+\tanh\left(\frac{x - x_c + w}{\sigma}\right)\right] \times \\ \times \left[1-\tanh \left(\frac{x - x_c - w}{\sigma}\right)\right]
\end{multline}
centered at {$x_c = 133\ \mu$m and of half-width $w = 74\ \mu$m}, and edge smoothening $\sigma = 10\ \mu$m. At the mean-field level considered in this section, the emitted signal is equal to the product of the in-cavity field $\phi(x,t)$ and the filter function $F(x)$.


In practice, in our simulations the field is described within a rotating frame at the pump frequency and the temporal dynamics given by the driven-dissipative Gross-Pitaevskii equation (\ref{eq:mf_gpe}) is followed until a steady state is reached. As the probe and pump frequencies $\omega_{s,p}$ are not identical, this steady state consists of a strong constant component corresponding to the pump, plus a weak oscillating modulation at the difference frequency. As a consequence of the nonlinearity of the GP dynamics, the total emission spectrum displays a four-wave mixed signal at $\omega_4=2\omega_p-\omega_s$ in addition to the pump and probe components at respectively $\omega_{p,s}$. In our numerics, a temporal Fourier transform can be used to isolate the $\omega_{s,4}$ frequency components from the stronger pump at $\omega_{p}$. Then, for each of the $\omega_{s,4}$ frequencies, a spatial Fourier transform of the spatially filtered field allows to separate the different wavevector components so to obtain the momentum distribution as usually done in a far-field measurement~\cite{Nguyen2015}. As an example, as it is pictorially explained in Fig.\ref{f:peaks}(b), a Bogoliubov excitation results in a finite signal in both frequency components $\omega_{s,4}$, peaked at wavevector values determined by the local Bogoliubov dispersion in the collection region, and weighted by the Bogoliubov $u^2$ and $v^2$ coefficients. 

While the finite size of the signal collection region and the intrinsic spatial decay of the scattered wave propagating away from the horizon are responsible for a broadening of the peaks in wavevector space, the spatial variation of the Bogoliubov dispersion due to the inhomogeneity of the condensate density and speed is negligible, as witnessed by the almost constant flow speed shown in Fig.~\ref{f:setup}. Note that an analogous measurement in the upstream region would instead be strongly disturbed by the strong inhomogeneity of the flow profile, as well as by the strong intensity of the underlying pump beam.

\subsection{The response signals}

Examples of the collected wavevector spectra at the two frequencies $\omega_s$ and $\omega_4$ are shown in Fig.\ref{f:peaks}(c-f) for different values of the detuning $\hbar\Delta=\hbar(\omega_s-\omega_p)=0.94,\,0.60,\,0.37,\,-0.60$\,meV, as indicated by the horizontal lines in the Bogoliubov dispersion at the probe position plotted in Fig.\ref{f:peaks}(a). The vertical {lines in Fig.\ref{f:peaks}(c-f)} indicate the wavevectors of the different scattered channels {as predicted} by the Bogoliubov dispersion at the collection point shown in Fig.\ref{f:peaks}(b) {: their position successfully compare with the peaks observed in the numerically computed spectra.}

For the highest value of the detuning $\Delta$ [panel (c)], the probe spectrum at $\omega_s$ shows a single peak due to a standard reflection process. This process also appears as a single peak in the four-wave mixed spectrum at $\omega_4$. The much weaker intensity of the latter is due to the almost pure single-particle nature of the high-energy Bogoliubov excitations considered here.

For decreasing $\Delta$, the signal from a Hawking mode-conversion process starts being visible. While at {$\omega_s$} it is often buried into the stronger signal from standard reflection, it is much more clearly visible at {$\omega_4$} where a doublet of peaks is apparent in panels (d,e). This enhanced visibility is a lucky combination of two factors: on one hand, the reflection signal is weaker at $\omega_4$ by the $u^2/v^2$ ratio of Bogoliubov coefficients; on the other hand, the same ratio for the Hawking wave enhances the feature in the $\omega_4$ signal. As $\Delta$ is further decreased, the Hawking and reflected features at $\omega_4$ approach each other, with the Hawking one eventually becoming the stronger one [panel (e)].

For negative $\Delta$, the physics is very similar albeit the exchanged position of the different spectral features [panel (f)]. Of course, for too large and negative $\Delta$, the Hawking feature disappears again (not shown).

For the sake of completeness, it is also interesting to look at this physics from the point of view of the spatial intensity profile of the $\omega_s$ and $\omega_4$ emissions. This is shown in Fig.~\ref{f:peaks}(g) for a detuning $\Delta\simeq 0.60$~meV as in Fig.\ref{f:peaks}(d). At $\omega_s$, there are fast oscillations in {the $0<x<x_s$ region between the probe location} and the horizon due to interference between the incident and the reflected waves. On the other hand, slower interference fringes due to the beating of the reflected and the Hawking features forming the doublet of peaks in Fig.\ref{f:peaks}(d-e) are clearly visible in the {$\omega_4$} signal {in this same region}.

\subsection{Hawking radiation spectrum}

\begin{figure}
\includegraphics[width=0.85\columnwidth]{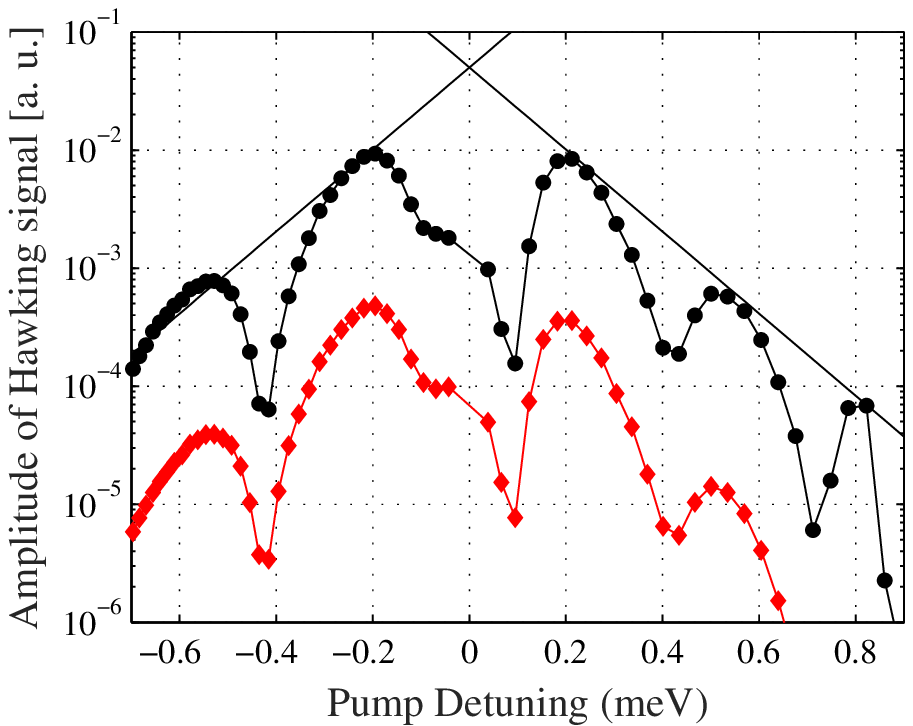}
\caption{
	Intensity of the stimulated Hawking fluorescence signal at the four-wave mixed frequency $\omega_4$ as a function of the probe detuning from the pump frequency $\hbar\Delta=\hbar(\omega_s-\omega_p)$. The Hawking signal is measured at the signal collection region (red) and is then corrected as to take into account velocity-dependent loss (black). Straight lines on either side of the distribution show the exponential envelope $I \propto \exp(-\hbar\,|\Delta|/k_B T)$ with a fitted Hawking temperature of $T=1.4\ $K. The strong periodic modulation is instead due to emergent resonant cavity.
}
\label{f:temperature}
\end{figure}

The series of wavevector spectra shown in Fig.\ref{f:peaks} are a good starting point to extract information on the frequency spectrum of the stimulated Hawking radiation. As the incident probe beam consists of relatively high-wavevector excitations on top of the condensate, we can reasonably neglect the dependence of its Bogoliubov coefficients on $\Delta$. As a result, the amplitude of the incident probe beam can be kept (approximately) constant while varying the probe wavevector $k_{s}$ by simply ensuring that the probe frequency $\omega_s$ and wavevector $k_s$ are always resonant with the desired point on the Bogoliubov dispersion. For each value of the probe frequency $\Delta$, the intensity of the stimulated Hawking radiation is estimated from the height of the Hawking peak in the four-wave mixed signal at $\omega_4$.

\subsubsection{Hawking temperature}

{As a first attempt to extract an estimate for the Hawking temperature of the black hole horizon, in Fig.\ref{f:temperature} we have plotted the probe frequency dependence of the height of the Hawking peak in the four-wave mixed signal at $\omega_4$. While the resulting spectrum shown as a red line displays the expected fast decay as a function of $|\omega_s|$, it also shows a unexpected marked asymmetry between the positive and negative $\Delta$ regions.}

As a possible explanation of this {unexpected feature that disturbs a unique determination of the Hawking temperature, we could} invoke the propagation losses between the probe injection point and the horizon, and then between the horizon and the collection region. While polariton losses give a constant temporal decay rate for Bogoliubov waves equal to $\gamma$, a significant frequency dependence can appear in the spatial absorption rate $\kappa=\gamma/v_g$ from the frequency-dependence of the group velocity $v_g$, which can be quite strong for the Hawking wave. In order to (at least partially) compensate for this spurious effect, a simplest strategy is to renormalize the observed signal in terms of the group velocity at the signal collection point,
\begin{equation}
 N_H=N_H^{\rm meas}(k)\,\exp\left[\gamma\left(\frac{L_1}{v_g(k_s)}+\frac{L_2}{v_g(k_H)}\right)\right]
\end{equation}
where $k_s$ is the wavevector of the incident probe wave and $k_H$ is the one of the scattered Hawking wave, and $L_{1,2}$ are the distances of the probe and the collection regions from the horizon. 

The renormalized data are shown in Fig.\ref{f:temperature} as a black line: the envelope is now well fitted on both $\Delta\gtrless 0$ sides by an exponential law with the same coefficient, which supports our interpretation of the asymmetry of the red curve of the bare data. Further renormalization taking into account other factors such as the frequency dependence of the Bogoliubov $u,v$ coefficients, the detailed spatial dependence of the group velocity, Jacobian coefficients due to the translation from wavevector to frequency, etc. would give corrections that scale at most as a power law of $\Delta$ and go beyond the precision of our analysis.

From the coefficient of the exponential law, one can extract an estimation for the Hawking temperature $T_H^{\rm est}=1.4$~K. In spite of the approximations underlying this estimation, this value is order-of-magnitude-consistent with the theoretical prediction based on the gravitational analogy \cite{Macher2009,Barcelo2011} 
\begin{align}
\label{eq:temperature}
T^{\rm th}_H &=  \left.\frac{\hbar}{2k_b c(x)} \frac{d}{dx}\left[v(x)^2-c(x)^2\right]\right|_{x=x_h} \simeq \nonumber \\
&\simeq \frac{\hbar}{2k_b c_h} \frac{(v_d^2-c_d^2)-(v_u^2-c_u^2)}{L_h} \simeq 3.5\,\mathrm{K},
\end{align}
where $L_h\sim 15\,\mu$m is the thickness of the horizon region where the speed of sound $c$ and flow velocity $v$ display their main variation, and subscripts $h,u,d$ stand for the `horizon', `upstream' and `downstream' regions in the close vicinity of the horizon. 

The fact that the estimated Hawking temperature $T_H^{\rm est}$ is somehow smaller than the analytical prediction $T_H^{\rm th}$ does not appear to be an issue, as this latter was derived in the hydrodynamic approximation where the flow and sound speeds vary slowly with respect to the healing length and, consequently, the Hawking temperature would be much smaller than the interaction energy $m c^2$. Inserting the actual values for the considered set-up, one finds a $T_H^{\rm th}/(mc_h^2)\simeq 2$ which indeed violates the hydrodynamic approximation. Based on the numerics of~\cite{Carusotto2008b} for atomic condensates, one expects that the actual Hawking temperature be somehow smaller than the analytical prediction $T_H^{\rm th}$, which confirms the consistency of our estimation.

Finally, it is worth highlighting that we are here comparing our numerical simulation for a driven-dissipative polariton fluid with the predictions of a gravitational analogy that was derived for standard particle-conserving fluids like atomic condensates: in the absence of theoretical results for the Hawking effect in driven-dissipative systems, our numerics suggest the remarkable conclusion that the thermality of the Hawking radiation process can be maintained also in an out-of-equilibrium context.

\subsubsection{Emergent resonant cavity}

\begin{figure*}
\includegraphics[width=0.85\textwidth]{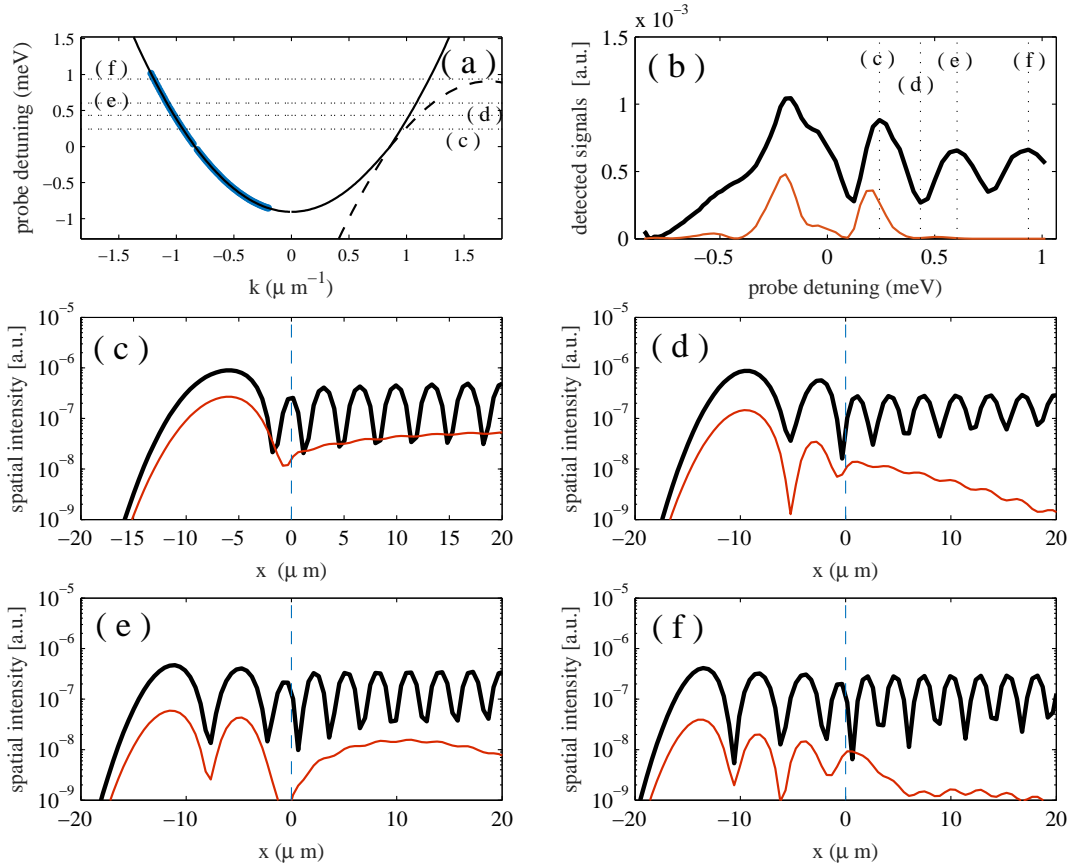}
\caption{Illustration of the resonance frequencies of the emergent cavity. {\bf Top-left (a) panel:} Bogoliubov dispersion relations at the spatial position of the probe. The horizontal lines mark the detunings values considered studied in the central and lower panels (c-f).
{\bf Top-right (b) panel:} {Total i}ntensity of the emitted signal at the probe (black) and four-wave mixed (red) frequencies $\omega_{s,4}$.
The former is due to a direct reflection of the probe excitations from the horizon. The latter includes contributions from both the four-wave-mixed component of the reflected beam and the stimulated Hawking emission. Vertical line marks the detunings of the (c-f) panels. {\bf Central and lower (c-f) panels:} spatial profile of the emission filtered at the probe frequency $\omega_{s}$ (solid black line) and at the four-wave mixed frequency $\omega_4$ (solid red line) for different probe detunings $\hbar\Delta \approx 0.25$~meV (c), $0.45$~meV (d), $0.61$~meV (e), $0.92$~meV (f). The vertical dashed line mark the approximate right bound of the emergent resonance cavity at the horizon location. The left boundary is located where the signal enters the high intensity area under the probe beam, namely around $x=-10\ \mu$m. The fringe pattern in the $\omega_s$ signal to the right of the horizon is due to interference of the incoming probe and the reflected wave. The peaks and dips in the four-wave mixed signal in panel (b) differ by half an oscillation period in the spatial patterns of (c-f). 
}
\label{f:fabry-perot}
\end{figure*}


In addition to the exponentially decaying envelope discussed in the previous sub-section, the Hawking spectrum plotted in Fig.\ref{f:temperature} exhibits a strong, almost periodic modulation. The underlying physics can be understood looking at the flow profile shown in the {lower} panel of Fig.\ref{f:setup} and, in particular, at the high-density region that is present at the pump position upstream of the defect {and that} is able to reflect all incident Bogoliubov excitations.

When the probe wave hits the horizon, it creates {reflected and Hawking waves} that propagate downstream, as well as a transmitted wave that propagates in the upstream region. When this last wave hits the high-density region {at the pump position}, it is completely reflected back towards the horizon. When {this} reflected wave hits the horizon, it can trigger a further stimulated Hawking process which also contributes to the Hawking wave. {It is the interference between these two contributions to the Hawking wave which is responsible for the marked oscillations that are visible in the Hawking spectrum of Fig.\ref{f:temperature} as well as for the spatial beatings that appear in the $0<x<x_s$ region between the probe location and the horizon in the spatial profiles of the $\omega_{s,4}$ emission shown in Fig.\ref{f:peaks}(g).}

Even though the strong inhomogeneity of the condensate prevents an easy analytical study of the scattering process, further insight on the interference process can be obtained from Fig.~\ref{f:fabry-perot}: the maxima/minima of the Hawking signal correspond to an integer/half-integer number of oscillations of the standing wave pattern in the $\omega_{s}$ intensity in the cavity region between the horizon and the pump region. In semi-quantitative terms, we can assume that the density (and hence the speed of sound) to be constant in this region: extracting from the flow profiles a speed of sound $c \approx 1.8\ \mu$m/ps, a length of the cavity $L \approx 10\ \mu$m, and using the fact that there must be an integer number of half-waves in the cavity for constructive interference, we get the energy level spacing between the levels $\hbar \Delta \omega = \pi \hbar c/L \approx 0.35\ $meV, which is in a fair agreement with the periodicity observed in the spectra of Fig.~\ref{f:fabry-perot}.

While resonant Hawking effect with a strongly modulated emission spectrum has been studied before by imposing an external cavity structure to the condensate flow~\cite{Zapata2011}, the most remarkable novelty of our finding stems from the fact that the cavity is not externally imposed, but naturally emerges as a by-product of the pumping scheme used to generate the black hole. Furthermore, our numerical simulations performed with different values of the pump intensity and different spatial positions of the pump spot show that the cavity structure is always present as long as we require the existence of a trans-sonic horizon.

\section{Spontaneous Hawking effect}\label{s:spon}

\begin{figure*}
\begin{tabular}{cc}
\includegraphics[width=0.85\columnwidth]{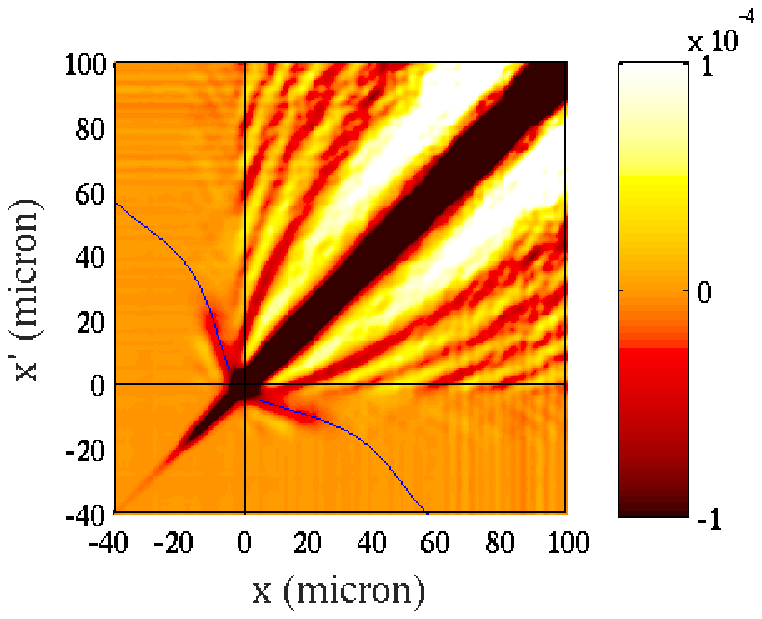} &
\includegraphics[width=0.85\columnwidth]{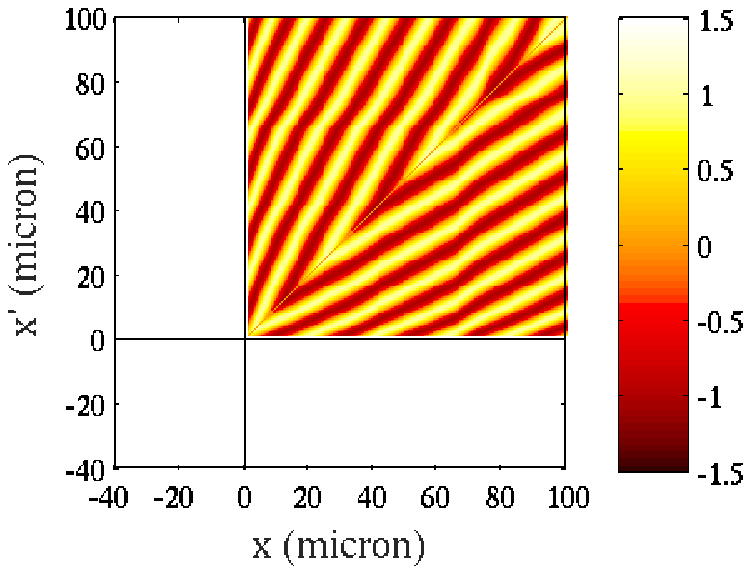} \\
\includegraphics[width=0.85\columnwidth]{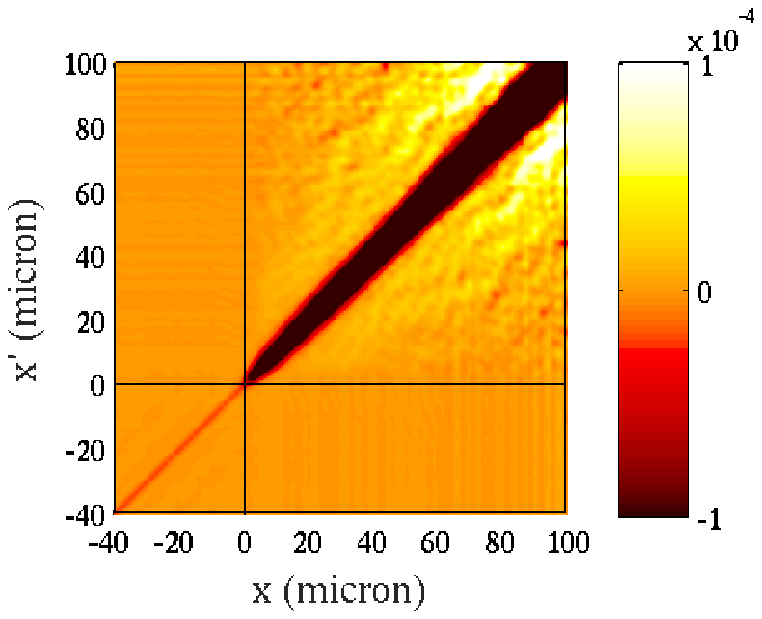} &
\parbox[b]{0.85\columnwidth}{\vspace{0.5cm}
\includegraphics[width=0.7\columnwidth]{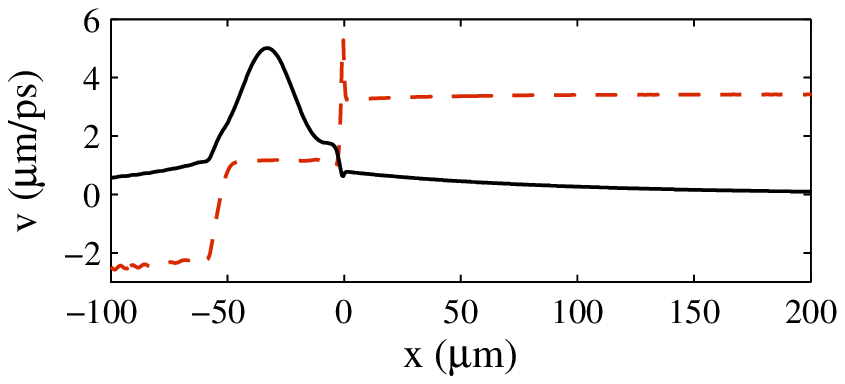} \\
\includegraphics[width=0.7\columnwidth]{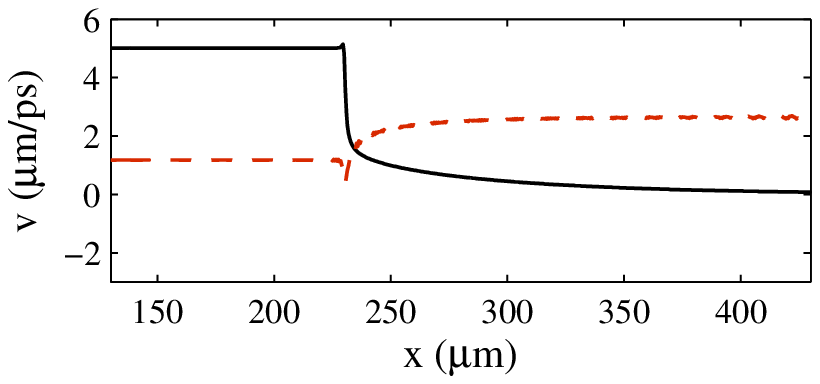}}
\end{tabular}
\caption{{\bf Top-left panel:} correlation function of density fluctuations showing various signatures of the spontaneous Hawking effect.  The blue solid line indicates the expected orientation of the Balbinot-Fabbri {correlation feature} between a Hawking phonon and its partner, emitted in the upstream and downstream directions, respectively. The prediction for the orientation is {obtained by inserting} the local values of the speed of sound and of the speed of flow in the vicinity of the horizon {into the differential equation Eq.\ref{e:locus}}. 
{\bf Top-right panel:} plot of the approximated analytical form (\ref{e:cos}) of the correlation signal between the Hawking partners emitted in the downstream direction at the specific frequency $\omega_{res} = 0.8\,$meV corresponding to the third resonance. In this simplified model, the correlation signal {has a $\cos[k x + k' x' +\varphi]$ form} in the $x'>x$ half-plane (and specularly inverted in the $x>x'$ half-plane), where the two wavevectors $k$,$k'$ are determined from inverting the dispersion relation $\omega(k) = \omega_{res}$.
 {\bf Bottom-left panel:} density correlation function for a {different flow configuration} where the horizon is very smooth and there is no emergent resonance cavity. As the surface gravity is very weak, no trace of Hawking processes is visible in the plot.
 {\bf Bottom-right panel:} Spatial profile of the flow velocity and of the speed of sound for the setup considered in the main text (top) and for a smooth horizon and no emergent resonant cavity (bottom).  
}
\label{f:twa}
\end{figure*}
In the previous Section we have concentrated our attention on a stimulated Hawking effect, where the mode conversion at the horizon is stimulated by a coherent probe beam of Bogoliubov excitations that scatter on the horizon. In the current section we proceed to analyze the spontaneous Hawking effect, where the same mode conversion process acts on the quantum vacuum of the Bogoliubov modes, so to convert the zero-point quantum fluctuations into correlated pairs of real Bogoliubov excitations propagating away from the horizon. As first proposed in~\cite{Balbinot2008a,Carusotto2008b} and widely discussed in the literature on analog models in cold atomic gases~\cite{Larre2012,Steinhauer2015} and polariton fluids~\cite{Gerace2012}, a most promising route to experimentally detect this spontaneous Hawking effect is through measurements of the intensity-intensity correlation function of the fluorescence signal.

This effect can be numerically studied by means of the so-called truncated Wigner approximation (TWA) \cite{Polkovnikov2010,Carusotto2013}, a semiclassical approach that allows to calculate the expectation values of quantum operators as classical averages of suitable stochastic partial differential equations. This method was originally introduced in the quantum fluids context for cold atom systems~\cite{Steel:PRA1998,Sinatra:JPhysB2002,Blakie2008} and soon extended to the driven-dissipative case of quantum fluids of light~\cite{Carusotto2005}. The basic idea of the TWA method is that one can {truncate the Fokker-Planck-like equation for the Wigner distribution of the quantum field $\hat{\psi}(x)$ to second-derivative terms, so to approximately map the quantum evolution of $\hat{\psi}(x)$ onto a stochastic differential equation for the corresponding classical field $\phi(x,t)$. 

On a numerical grid of spacing $\Delta x$, the resulting stochastic Gross-Pitaevskii equation for our specific system} reads
\begin{align}
\label{eq:wigner}\nonumber
i \, d\phi &=\left[- \frac{\hbar}{2m} \frac{d^2}{dx^2}
+ V(x) + g\left(|\phi |^2 - \frac{1}{\Delta x}\right)  - i \frac{\gamma}{2}  \right]  \phi \,dt +
\\
&+ F_{p}(x,t) \, dt 
+  \sqrt{\frac{\gamma}{4}} \,d \xi  \, ,
\end{align}
where $d \xi(x,t)$ is a complex, zero-mean, random-phase, Gaussian noise term with the correlator
\begin{equation}
\langle d \xi^*(x,t)\,d \xi(x',t')\rangle=\frac{2\, dt}{\Delta V} \, \delta_{x,x'}\, \delta ({t}-{t}').
\end{equation}
As long as interaction constant is weak enough for {the errors ensuing from the truncation approximation} to be negligible~\cite{Polkovnikov2010,Carusotto2013}, the classical averages $\langle \ldots \rangle_W$ over a large number of independent realizations of the classical field $\phi(x,t)$ then provide a faithful estimation of the corresponding equal-time, symmetrically-ordered observables. After converting to normally-ordered quantities, one obtains for the spatial density
\begin{equation}
\langle \hat\psi^\dag(x) \hat\psi(x)\rangle = \langle |\phi(x)|^2\rangle_W - \frac{1}{2\Delta x},
\label{eq:stoch_dens}
\end{equation}
and for its fluctuations
\begin{multline}
\langle \hat\psi^\dag(x)  \hat\psi^\dag(x') \hat\psi(x') \hat\psi(x)\rangle = \\
=\langle |\phi(x)|^2\, |\phi(x')|^2\rangle_W + 
 \frac{1}{4 \Delta x^2}(1 - \delta_{x,x'}) +
\\
- \frac{1}{2\Delta x} (1 - \delta_{x,x'})
\langle |\phi(x)|^2 + |\phi(x')|^2\rangle_W.
\label{eq:stoch_correl}
\end{multline}

As we are interested in the steady-state observables under a single monochromatic pump at $\omega_p$ (no probe is needed to study the spontaneous Hawking effect), in our numerics we simulate the temporal evolution given by stochastic differential equation (\ref{eq:wigner}) for long times. Once a steady-state is reached, the stochastic field is repeatedly sampled at periodic intervals: from this sampling, statistical estimates of the quantum observables are obtained by means of the above-mentioned correspondence of classical averages over stochastic noise and quantum expectation values, see e.g. (\ref{eq:stoch_dens}) and (\ref{eq:stoch_correl}). Provided the temporal spacing of the samples is long enough for them to be statistically independent, the statistical error decreases with the number $N_{\rm sam}$ of samples as $1/\sqrt{N_{\rm sam}}$.

As a main outcome of our calculations, numerical results for the normalized correlation function of density fluctuations defined as
\begin{equation}
g^{(2)}(x,x') = \frac{\langle \hat{\psi}^{\dagger} (x) \hat{\psi}^{\dagger} (x') \hat{\psi}(x') \hat{\psi}(x)  \rangle}{
\langle \hat{\psi}^{\dagger} (x) \hat{\psi}(x)  \rangle
\langle \hat{\psi}^{\dagger} (x') \hat{\psi}(x')  \rangle},
\end{equation}
are shown in the top-left panel of Fig.~\ref{f:twa} for the same pump configuration considered in the previous sections of this work.

While the negative correlation signal along the main diagonal is a standard many-body effect stemming {due to the} repulsive interactions in the fluid {and has little to do with the horizon~\cite{Carusotto2008b}}, the most evident feature of the Hawking radiation is the off-diagonal negative correlation signal {that encodes the correlation within the excitation pairs} that are simultaneously emitted in the Hawking process: one {excitation} escapes from the black hole in the negative $x<0$ direction, while the partner {one} propagates in the opposite direction $x>0$ falling into the black hole. 
In contrast to atomic systems where it keeps growing in length for indefinite times after formation of the horizon~\cite{Balbinot2008a,Carusotto2008b}, this Balbinot-Fabbri correlation feature has a finite steady-state length in the present driven-dissipative case due to two reasons: on one hand, the finite lifetime of polaritons limits the propagation distance of the Hawking excitations to a distance $v_g/\gamma$ on the order of few tens of microns on either side of the horizon. On the other hand, the high-density region present right upstream of the horizon reflects the Hawking emission and distorts the correlation signal.

In addition to this, the spatial inhomogeneity of the fluid is also responsible for a curved shape of the Balbinot-Fabbri feature, as also noticed in~\cite{Nguyen2015}: the {blue} line in the top-left panel of the Fig.~\ref{f:twa} shows the locus of points of equal optical distance from the horizon. Neglecting dispersion, this is locally defined (in the $x>0$, $x'<0$ sector) by the differential equation
\begin{equation}
 \frac{dx}{v_g(x)}= \frac{dx'}{v_g(x')}
 \label{e:locus}
\end{equation}
where the group velocity $v_g$ is evaluated at each point for the relevant Bogoliubov branch. 
As one can see {from} the flow profile displayed in the lower-right panel, the curvature of the locus is mostly due to the variation in the speed of sound in the $x<0$ region. In addition to the curvature, one can also notice additional fringes parallel to the main axis of the feature, in particular a weakly positive signal at lower $x<0$: the presence of these fringes can be physically explained by the frequency-dependence of the group velocity of Bogoliubov excitations, mostly of the Hawking branch~\cite{Recati2009}.

Another feature which directly relates to the Hawking effect is the positive correlation stripe located parallel to the main diagonal in the $x,x'>0$ downstream region, followed by analogous, yet weaker fringes further away from the main diagonal. This complex feature was already present in the previous work~\cite{Nguyen2015}, but no physical explanation was given for its structure, markedly different from the moving fringes found in the same spatial region in~\cite{Carusotto2008b} and indicated as feature (ii): in the present calculations, the fringe pattern {is} in fact a steady-state feature which does not drift with time. Furthermore, in contrast to feature (iv) in~\cite{Carusotto2008b}, it shows clear oscillations.

Our explanation for this feature goes back to the frequency-dependence of the Hawking emission in the downstream direction: making use of the usual quantum optical criterion~\cite{Liscidini} to translate the stimulated Hawking spectrum discussed in Sec.\ref{f:temperature} to spontaneous processes, one expects that the emergent cavity modulates the spontaneous Hawking emission spectrum into a series of narrow peaks at which the emission intensity is concentrated. 

As the emission in this downstream region consists of correlated pairs of Bogoliubov excitations at opposite frequencies, we can expect that the time-independent anomalous correlation $m_{k,k'} = \langle a_k a_{k'} \rangle$ between right-propagating modes of {opposite frequencies} can give rise to a peculiar interference feature in the $g^{(2)}(x,x')$ intensity correlation function. {Indicating with $k,k'$ the wavevectors of these modes relative to the flowing condensate, the interference feature can be shown to have the form
\begin{equation}
(u_{k'}+v_{k'})(u_k+v_k) \left[m_{k,k'} e^{i(k x + k' x')} +\textrm{c.c.}\right]
\label{e:cos}
\end{equation}}
in the $x'>x$ half-plane and a specularly inverted one in the $x>x'$ half-plane. Here, the $u_k+v_k$ and $u_{k'}+v_{k'}$ factors quantify the density component of the Bogoliubov modes. When integrating over all pairs of modes, it is natural that the largest contribution will be provided by the peaks in the Hawking spectrum, which result in weakly damped spatial oscillations.

For illustrative purposes, we focus on the third peak in the Hawking signal at $\omega_{res} \approx 0.8\ $meV [labelled (f) in Fig.\ref{f:fabry-perot}(b)]. For this peak, the wavevectors of the two right-propagating modes at frequencies $\pm \omega_{res}$ are then at $k \approx 0.3\ \mu$m$^{-1}$ and $k' \approx -0.6\, \mu$m$^{-1}$ {relative to the flowing condensate}. The shape of the resulting signal~\footnote{The association of the $k,k'$ modes with the $x,x'>0$ position in the expression (\ref{e:cos}) must be made in terms of the group velocity of the modes $v_g(k),v_g(k')>0$: on physical grounds, the more remote point has to be associated to the faster group velocity.}  is plotted in the top-right panel of Fig.~\ref{f:twa}: the qualitative agreement with the period and the orientation of outer fringes that are visible around $x=20\,\mu$m and $x'=80\,\mu$m in the full numerical calculation shown in the top-left panel is qualitatively quite good {and supports} our interpretation. 

Closer to the main diagonal, the simultaneous presence of other processes makes the signal more complicated. For instance, an analogous contribution from the lower peaks in the Hawking spectrum is responsible for the wider and positive first fringe (around $x=60\,\mu$m and $x'=80\,\mu$m): further away from the diagonal, the relative importance of these contributions is however suppressed {by the wider linewidth} of the lower Hawking peaks which gives a quicker decay of the corresponding fringes.

As a final check of our interpretation, in the lower-left panel we have plotted the same $g^{(2)}$ correlation function for a different flow configuration where the surface gravity is very low. This configuration is obtained with a spatially homogeneous pump restricted to the left of the defect (a sort of theta-function pump) and gives rise to the flow and sound speed profiles displayed in the two lower-right panels. As expected, all interesting features disappear exception made for the negative correlation signal along the main diagonal which is due to many-body effects and has no relation to the horizon: in particular, {there are no fringes} in the $g^{(2)}$ nor any Hawking signal. 


\section{Conclusions}\label{s:concl}

In this work we have proposed and numerically characterized an experimental setup to study the Hawking effect in a flowing fluid of exciton-polaritons in a laterally patterned semiconductor microcavity device under a coherent monochromatic pump.
A stimulated analog Hawking effect can be studied in a pump-and-probe-type measurement, while the spontaneous analog Hawking effect is observable in the correlations function of the intensity noise in the secondary emission. The main conclusion of this work is that both effects can be observed in a state-of-the-art device with standard quantum optical tools.

In order to detect the stimulated Hawking emission, we have proposed to shine an additional weak and monochromatic probe beam onto the cavity, so to generate a coherent Bogoliubov excitation propagating against the horizon. The stimulated Hawking emission is detected by isolating the scattered waves at the probe and four-wave-mixed frequencies and measuring their wavevector distribution. With respect to previous studies of wavepacket scattering on a black hole horizon~\cite{Gerace2012}, the pump-probe experiment proposed here appears to be technologically much easier to implement as it only requires a pair of continuous-wave laser beams and a angularly- and spectrally-selective detection system, with no need for time-resolved technology to generate and detect the time-dependent signals of a pulsed laser. 

In particular, the numerically observed exponential decay of the scattered amplitude at the four-wave-mixed frequency as the function of the pump-probe detuning provides a prediction for the Hawking temperature around $1.4\ $K in good agreement with the value expected from the surface gravity of the horizon. 
As a novel unexpected phenomenon, we identified the spontaneous appearance of an emergent resonant Fabry-Perot-type cavity for sound waves, formed by the strong pump beam, which completely reflects the quasiparticles, and the semi-transparent horizon. As a consequence of this emergent resonant cavity, the stimulated Hawking response shows a strong modulation on top of the exponential decay, with well-defined peaks at the cavity resonances.

We have finally analyzed the correlations in the intensity noise of the secondary emission for the same setup in the absence of the probe beam and showed that the spontaneous Hawking effect gives rise to peculiar correlation features: in addition to the moustache-shaped Balbinot-Fabbri correlation signal between the inner and the outer regions of the black hole, we have identified a regular fringe pattern in the inner region and we have interpreted it as a consequence of the marked frequency-dependence of the Hawking emission.

The natural next step of our investigation is of course to obtain an experimental verification of our predictions, first for the stimulated and then for the spontaneous Hawking emission. From the theoretical point of view, we are presently working on the extension of our theory to deal with the quantum entanglement features that between the two sides of the horizon because of the spontaneous Hawking processes: as compared to existing works on this physics in the case of atomic condensates~\cite{DeNova2012,Finazzi2013,Steinhauer2015}, we expect that the intrinsically driven-dissipative nature of the photon/polariton fluid will be responsible for significant differences in the quantum dynamics and, in particular, a most challenging task will be to identify schemes able to extract an entanglement signal out of the extra noise due to dissipation~\cite{Busch2014}. 

Another most intriguing next step will be to theoretically investigate the potential of our setup to simulate the so-called Hartle-Hawking vacuum of an eternal black hole \cite{Birrell1982,Hartle1976}. The eternal black hole can be thought of as a conventional black hole, which is in thermal equilibrium with infalling radiation. One way to achieve this state is to put a black hole in the center of a perfectly reflecting spherical shell. The shell will reflect the emitted Hawking radiation back, so eventually the black will equilibrate at the same temperature as the ambient thermal field. In our system, the reflecting shell naturally appears as a consequence of the high-intensity pump region which reflects quasiparticles back to the horizon. With respect to the presently available configurations with a relatively short cavity and well-separated resonances, the challenge will be to obtain a sufficiently long cavity for which the acoustic fluctuations between the pump and the horizon may end up in a thermal state with the Hawking temperature.

\section*{Acknowledgements}
We are thankful to P.-\'E.~Larr\'e and I.~Khavkine for stimulating discussions and to S.~Finazzi for pointing out the renormalization of the Hawking temperature due to velocity-dependent loss. This work was supported by the {EU-FET Proactive grant AQuS, Project No. 640800, by the ERC grants Honeypol and QGBE, and by the Autonomous Province of Trento, partly under the call
``Grandi Progetti 2012'' project ``On silicon chip
quantum optics for quantum computing and secure
communications--SiQuro''.}

\bibliography{library}

\end{document}